\documentclass[12pt]{iopart}
\usepackage{graphicx}% Include figure files
\begin{document}

\title[Superconductivity and ferromagnetism in EuFe$_{2}$(As$_{1-x}$P$_{x}$)$_{2}$]
{Superconductivity and ferromagnetism in
EuFe$_{2}$(As$_{1-x}$P$_{x}$)$_{2}$\footnote[1]{The main result of
this paper was presented in the 12th National Conference on Low
Temperature Physics held in July 2009, and Hangzhou Workshop on
Quantum Matter held in October 2009.}}

\author
{Guanghan Cao, Shenggao Xu, Zhi Ren \footnote[2]{Present Address:
Institute of Scientific and Industrial Research, Osaka University,
Ibaraki, Osaka 567-0047, Japan}, Shuai Jiang \footnote[3]{Present
Address: I. Physikalisches Institut, Georg-August-Universit\"{a}t
G\"{o}ttingen, D-37077 G\"{o}ttingen, Germany}, Chunmu Feng, Zhu'an
Xu}

\address{State Key Lab of Silicon Materials and Department of Physics, Zhejiang University, Hangzhou 310027, China}

\ead{ghcao@zju.edu.cn}

\begin{abstract}
Superconductivity and ferromagnetism are two antagonistic
cooperative phenomena, which makes it difficult for them to coexist.
Here we demonstrate experimentally that they do coexist in
EuFe$_{2}$(As$_{1-x}$P$_{x}$)$_{2}$ with $0.2\leq x\leq0.4$, in
which superconductivity is associated with Fe-3$d$ electrons and
ferromagnetism comes from the long-range ordering of Eu-4$f$ moments
via Ruderman-Kittel-Kasuya-Yosida (RKKY) interactions. The
coexistence is featured by large saturated ferromagnetic moments,
high and comparable superconducting and magnetic transition
temperatures, and broad coexistence ranges in temperature and field.
We ascribe this unusual phenomenon to the robustness of
superconductivity as well as the multi-orbital characters of iron
pnictides.

\ (Some figures in this article are in colour only in the electronic
version)

\end{abstract}


\submitto{\JPCM} a special issue on New Trends in Superconducting
Materials


\maketitle

\section{Introduction}
Since the discovery of superconductivity (SC) in mercury 100 years
ago,\cite{onnes} thousands of superconductors have been found in
various kind of materials including elements, alloys, inorganic
compounds and organic compounds. Nevertheless, SC was shown to be
incompatible with magnetism, simply manifested by that fact that all
the magnetic elements do not superconduct. Although SC may coexists
with antiferromagnetism because superconducting coherent lengths are
generally much larger than interatomic distances, it is easily
destroyed by ferromagnetism (FM) owing to orbital\cite{ginzburg} and
paramagnetic\cite{cc62} effects. On the other hand, SC does not
support the local-moment FM via Ruderman-Kittel-Kasuya-Yosida (RKKY)
interactions.\cite{anderson59} Earlier experiments revealed the
incompatible nature of the two collective phenomena in (La,Gd) and
(Ce, Pr)Ru$_2$ solid solutions.\cite{matthais} Until late 1970s
possible coexistence of SC and FM was evidenced in
ErRh$_4$B$_4$\cite{fertig77} and
Ho$_{1.2}$Mo$_6$S$_8$\cite{ishikawa}, only under narrow regimes of
temperature and external field. The interplay of SC and magnetic
ordering was also exhibited in a family of layered compounds
$R$Ni$_2$B$_2$C ($R$=Tm, Er, Ho and Dy)\cite{rev1221} and
rutheno-cuprates\cite{felner,bernhard}. Another interesting
coexistence was found in UGe$_2$\cite{UGe} and URhGe,\cite{URhGe} in
which SC occurs under the ferromagnetic background ($T_c<T_M$) and,
both SC and FM come from the same type of electrons.

The discovery of Fe-based superconductors\cite{hosono, rev2010}
brought about new findings on the interplay of SC and FM.
EuFe$_2$As$_2$, first synthesized in late 1970s,\cite{Eu122} is a
unique "122" compound which shows both SC in the FeAs-layers upon
appropriate doping, and long-range magnetic ordering in the Eu
sublattice. It was found that the undoped parent compound undergoes
antiferromagnetic (AFM) ordering in the Fe sublattice at 200 K,
followed by another AFM ordering in the Eu sublattice at 20
K.\cite{Raffius,Ren2008,JOP2008,Jeevan1} The two subsystems are
hardly coupled, as evidenced from optical\cite{wu} and
photoemission\cite{fdl} studies. The magnetic structure of the
latter AFM order had been proposed to be of
$A$-type,\cite{Ren2008,Jiang} in which Eu$^{2+}$ spins algin
ferromagnetically in the basal planes but antiferromagnetically
along the $c$-axis, which was then confirmed by the magnetic
resonant x-ray scattering\cite{RXS} and neutron diffraction\cite{ND}
experiments. By the partial substitution of Eu with K, SC over 30 K
was reported in Eu$_{1-x}$K$_{x}$Fe$_2$As$_2$.\cite{Jeevan2}
However, due to the dilution effect by the Eu-site doping, no
magnetic ordering for Eu$^{2+}$ spins was observed. In attempt to
obtain SC by Ni doping in EuFe$_{2-x}$Ni$_x$As$_2$\cite{Ren-Ni}, we
observed FM ordering for the Eu$^{2+}$ moments. In the case of Co
doping, however, there was a superconducting transition at $\sim$21
K, followed by resistivity reentrance around 17
K.\cite{Eu122Co,cxh-Co} By P doping at the As-site, we found both SC
at $T_c$=26 K and local-moment FM in the Eu sublattice at 20 K in
EuFe$_{2}$(As$_{0.7}$P$_{0.3}$)$_{2}$.\cite{Ren-P} In this paper we
report schematic investigations on how SC and FM are evolved in
EuFe$_2$(As$_{1-x}$P$_x$)$_2$ system, particularly focusing on where
and why SC coexists with FM.

\section{Experimental}

Polycrystalline samples of EuFe$_2$(As$_{1-x}$P$_x$)$_2$ ($x$=0,
0.05, 0.10, 0.15, 0.20, 0.25, 0.30, 0.35, 0.40, 0.45, 0.50, 0.75 and
1.0) were synthesized by solid state reaction with EuAs, Fe$_{2}$As
and Fe$_{2}$P. EuAs was prepared by reacting fresh Eu grains and As
powders in evacuated quartz tube at 873 K for 10 h then 1023 K for
another 10 h, and ultimately 1223 K for 10 h. Fe$_{2}$As was
presynthesized by reacting Fe powers and As powders at 873 K for 10
h and 1173 K for 15 h. Fe$_{2}$P was prepared by heating Fe powders
and P powders very slowly to 873 K and then holding for 10 h. In an
argon-filled glove-box, the powders of EuAs, Fe$_{2}$As and
Fe$_{2}$P were mixed in a certain stoichiometric ratio, thoroughly
ground in an agate motar and pressed into pellets. The pellets were
annealed in an evacuated silica tube at 1273 K for 20 h and
furnace-cooled to room temperature. The resultant
EuFe$_2$(As$_{1-x}$P$_x$)$_2$  samples were black in colour and
rather stable in air.

Powder x-ray diffraction (XRD) was performed at room temperature,
using a D/Max-rA diffractometer with Cu-K$_{\alpha}$ radiations and
a graphite monochromator. The data were collected with a step-scan
mode for $10\,^{\circ}$$\leq$$2\theta$$\leq$$80\,^{\circ}$. Lattice
parameters were refined by a least-squares fit with considerations
of zero shift. The electrical resistivity was measured using a
standard four-probe method with the applied current density of
$\sim$0.5A/cm$^{2}$. The dc magnetization was measured on a Quantum
Design Magnetic Property Measurement System (MPMS-5).

\section{Results and discussion}

Figure 1(a) shows the XRD patterns of the
EuFe$_2$(As$_{1-x}$P$_x$)$_2$ samples. The specimen are basically of
single phase with ThCr$_{2}$Si$_{2}$-type structure. For some
samples, only minor amount ($<5\%$) of impurities of Eu$_{2}$O$_{3}$
and Fe$_{2}$P was detected. The XRD patterns show systematic changes
with the P content. For example, the intensity of (002) peaks
increases gradually with incorporating phosphorus. Both (008) and
(200) peaks shift towards higher angles with increasing $x$,
indicating the lattice shrink. Indeed, we see that both $a$ and $c$
decrease with $x$ in figure 1(b). Notably, the magnitude of decrease
in $c$ is much bigger. This is very important for the evolution of
magnetic ordering in Eu sublattice, since the RKKY coupling is
closely related to the interatomic distance of the magnetic
ions\cite{Ren-Ni}.

\begin{figure}
\center
\includegraphics[width=15cm]{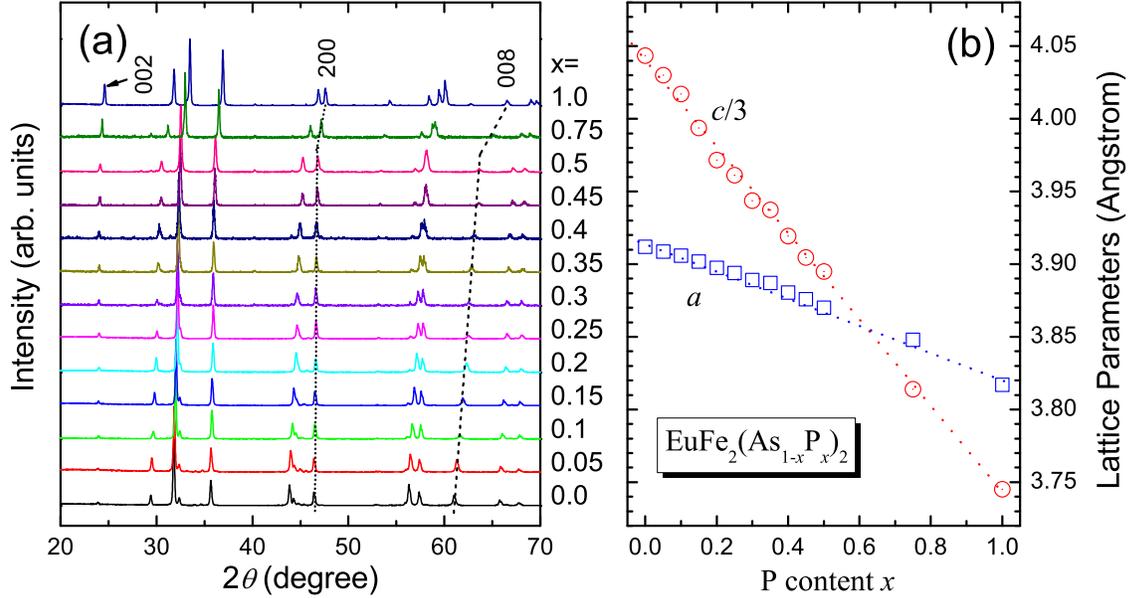}
\caption{(a) Powder X-ray diffraction patterns of
EuFe$_{2}$(As$_{1-x}$P$_x$)$_2$ samples at room temperature. (b)
Refined lattice parameters plotted as functions of P content $x$.}
\end{figure}

Figure 2 shows the temperature dependence of resistivity for
EuFe$_{2}$(As$_{1-x}$P$_{x}$)$_{2}$. The undoped EuFe$_{2}$As$_{2}$
clearly shows two resistivity kinks at 200 K and 20 K, associated
with Fe-site and Eu-site AFM ordering, respectively. Note that the
transition temperatures for single crystal samples are 190 K and 19
K, respectively.\cite{wu} This slight difference is demonstrated to
be due to tiny Eu-deficiency in single crystal samples (In this
sense the chemical stoichiometry can be better kept using
polycrystalline samples). With the P doping, the magnetic transition
temperature of Fe sublattice ($T_{M}^{Fe}$) decreases rapidly,
similar to the case in
BaFe$_{2}$(As$_{1-x}$P$_{x}$)$_{2}$\cite{Ba122P}. On the other hand,
the magnetic transition temperature of Eu sublattice ($T_{M}^{Eu}$)
first decreases by 4 K, then starts to increase at $x\sim$0.15. When
the P content increases up to 20\%, no anomaly associated with the
Fe-AFM ordering can be observed, instead, a sudden decrease in
resistivity is seen below 21 K, indicative of superconducting
transition. Besides, a shoulder (resistivity reentrance) appears at
16 K, which is probably due to the magnetic ordering in Eu
sublattice. The optimal doping is at $x$=0.3 where the onset
superconducting transition temperature ($T_{c}^{onset}$) achieves
the maximum (29 K) with no resistivity reentrance at lower
temperatures for our best sample. When $x>$0.45, no sign of SC was
detected, and the resistivity kinks reappear at $T_{M}^{Eu}$. As can
be seen, $T_{M}^{Eu}$ climbs with further P substitution, and
arrives at 29 K for the end member EuFe$_2$P$_2$\cite{feng,ryan}.

\begin{figure}
\center
\includegraphics[width=15cm]{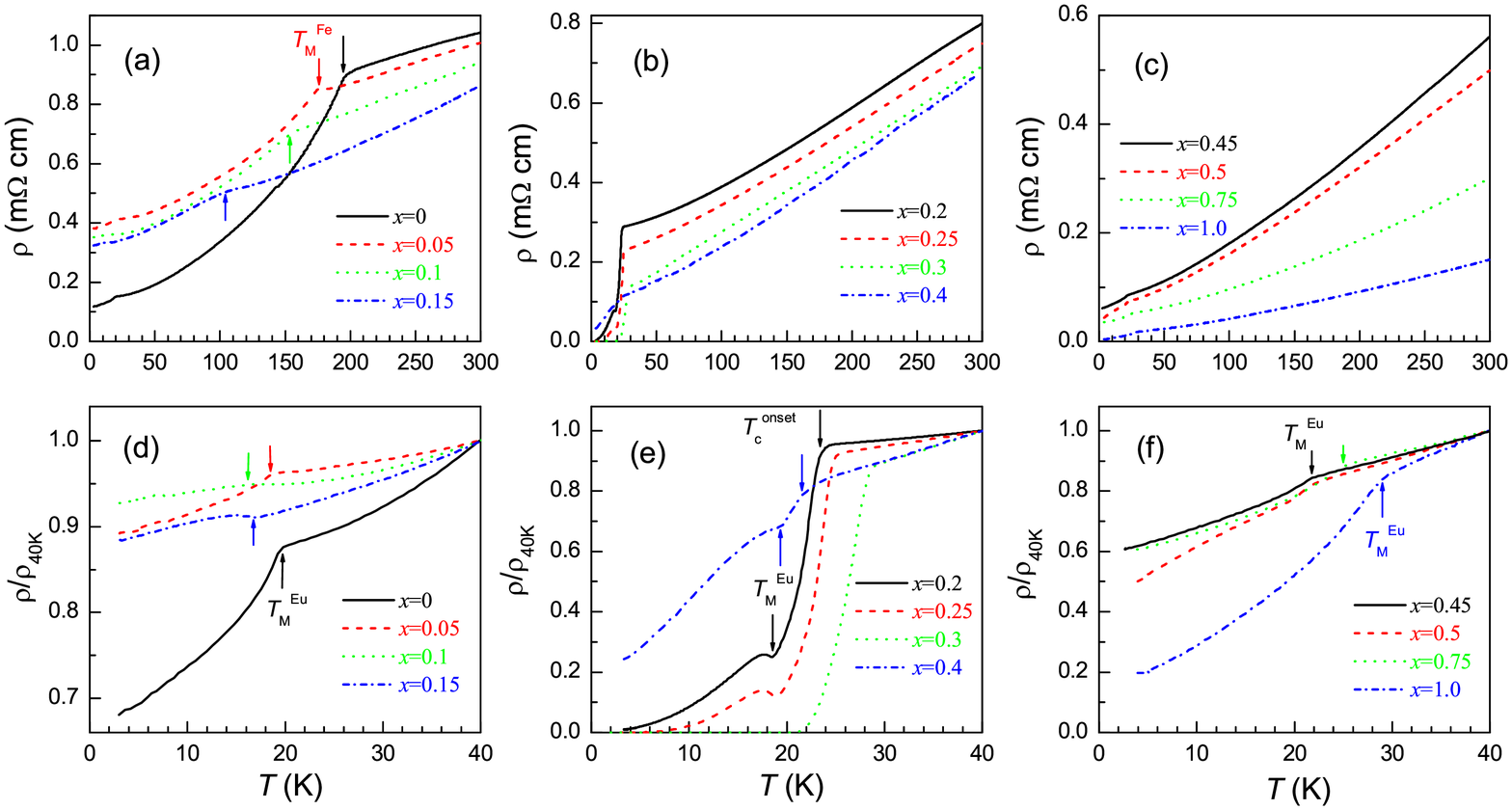}
\caption{Temperature dependence of resistivity for
EuFe$_{2}$(As$_{1-x}$P$_x$)$_2$ polycrystalline samples. To
explicitly show the regimes that have different characteristic, we
separate into three groups from right to left, corresponding to
lower, intermediate and higher doping. The lower panels display the
expanded plots using normalized resistivity for comparison. The
arrows mark the positions of the related transition temperatures.}
\end{figure}

Figure 3 shows the temperature dependence of magnetic susceptibility
of two representative samples. For $x$=0.05, the $\chi (T)$ curve
has a peak at $T_{M}^{Eu}$=18.5 K. Meanwhile, there is bifurcation
for FC and ZFC branches, suggesting ferromagnetic component. The
ferromagnetic component can be understood by recent M\"{o}ssbauer
study\cite{felner-jop} which shows canting of the Eu spins toward
the $c$-axis with P doping. The spin canting was very recently
evidenced by measuring the anisotropic magnetization for
EuFe$_{2}$(As$_{0.88}$P$_{0.12}$)$_2$ single crystals.\cite{zapf}
There is another anomaly at $\sim$6 K, suggesting more complicated
successive magnetic transition.

For $x$=0.25, the FC curve looks like the behaviour of a typical
ferromagnet. The low-temperature susceptibility is 10 times larger
than that of $x$=0.05. More obvious bifurcation is seen for the ZFC
and FC curves. The $M-H$ loop (not shown here) has a clear magnetic
hysteresis, like that previously reported for $x$=0.3\cite{Ren-P}.
The saturated magnetization value corresponds to the fully-aligned
Eu moments (see below). All these facts indicate ferromagnetic
ordering for the Eu$^{2+}$ spins. The ferromagnetic Curie
temperature is identified at $T_{M}^{Eu}$=19 K, above which the
$\chi (T)$ data obey extended Curie's Law
($\chi=\chi_{0}+C/(T-\theta)$, where $\chi_0$ denotes the
temperature-independent term, $C$ the Curie-Weiss constant and
$\theta$ the paramagnetic Curie temperature). The data fitting gives
the effective moment for Eu, $P_{eff}$=8.3 $\mu_{B}$ per formula
unit, which is close to the theoretical value
$g\sqrt{S(S+1)}\mu_{B}$=7.94 $\mu_{B}$ ($S=7/2$ and $g$=2) for a
free Eu$^{2+}$ ion. The $\theta$ values are positive, in agreement
with the dominant ferromagnetic interaction among Eu$^{2+}$ spins.

\begin{figure}
\center
\includegraphics[width=15cm]{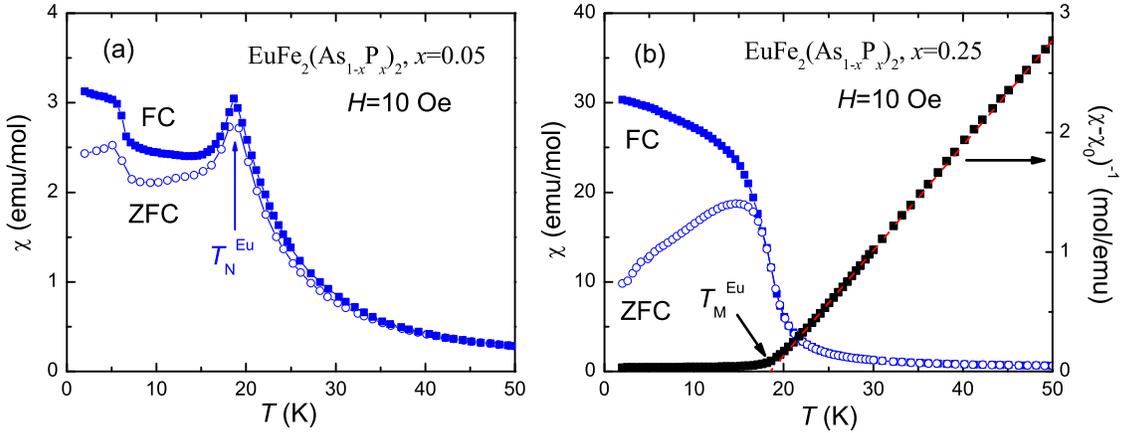}
\caption{Temperature dependence of magnetic susceptibility for
EuFe$_{2}$(As$_{1-x}$P$_x$)$_2$. (a) $x$=0.05; (b) $x$=0.25.}
\end{figure}

Figures 4 displays the field dependence of magnetization for the
$x$=0.3 sample. Under the field of $\sim$1 T, the magnetization
saturates to 7.1 $\mu_B$/f.u. at 5 K. The saturated magnetization is
very close to the theoretical value of $gJ$=7.0 $\mu_B$/f.u.,
indicating fully polarization of the Eu$^{2+}$ spins. This fact
further confirms the ferromagnetism. The non-linear $M-H$ relations
above $T_{M}^{Eu}$ (20 K and 23 K) reflects that the external fields
help to align the Eu spins ferromagnetically.

The low-field magnetization [figure 4(b)] does not show magnetic
repulsion and shielding, as expected for a superconducting state.
The absence of Meissner effect could be due to the formation of
spontaneous vortex (SV) phase\cite{ng}. On the other hand, the loss
of magnetic shielding means that the resistivity is not really zero,
which is ascribed to the weak links in polycrystalline samples
and/or, the motion of SV which generates flow resistance.
Nevertheless, signature of SC can be seen from the concave curvature
of the $M-H$ relation at 23 K (which is higher than $T_{M}^{Eu}$ but
lower than $T_{c}^{onset}$), where diamagnetism is inferred from the
background of paramagnetism of Eu$^{2+}$ spins.

\begin{figure}
\center
\includegraphics[width=15cm]{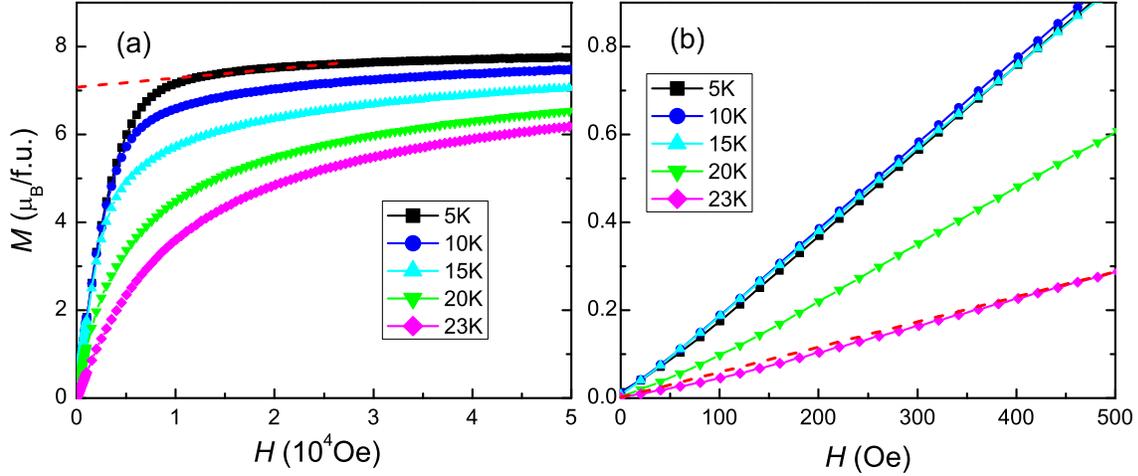}
\caption{Field dependence of magnetization at different temperatures
for EuFe$_{2}$(As$_{0.7}$P$_{0.3}$)$_2$ polycrystalline sample.}
\end{figure}

Based on the above results, we establish the magnetic and
superconducting phase diagram as shown in figure 5. For the Fe
sublattice, the phase diagram is similar to those of other Fe-based
superconductors\cite{Ba122P,cxh}, i.e., the Fe-AFM is suppressed by
the P doping, and then SC emerges. The only difference is that SC is
no longer alive when $T_{c}<T_{M}^{Eu}$. For the Eu sublattice, the
parent EuFe$_{2}$As$_{2}$ is $A$-type AFM ordered with Eu spins
lying along the $a$-axis. Doping with phosphorus not only increases
the interlayer RKKY coupling, but also leads to the canting of the
Eu spins. According to the M\"{o}ssbauer study,\cite{felner-jop} the
spin canting starts at $x\sim$0, and finishes at $x\sim$ 0.2, where
the spin-canting angle is $\sim 20^{\circ}$ from the $c$-axis.
Therefore, for $0.2\leq x\leq0.4$, SC definitely coexists with FM
even at zero field. Our recent magnetic Compton scattering
experiment for EuFe$_{2}$(As$_{0.73}$P$_{0.27}$)$_2$ also confirms
this scenario.\cite{deb} For $x>0.4$, only Eu-FM state is shown.
Here we should mention a recent result on the phase diagram, which
positions the SC dome in a narrower window around
$x$=0.2.\cite{Jeevan3} This discrepancy may be due to the
Eu-deficiency in the single crystal samples. Furthermore, it was
suggested that SC only coexists with Eu-AFM. However, their
following work reconciled the discrepancy in terms of the canting of
Eu$^{2+}$ spins.\cite{zapf}

\begin{figure}
\center
\includegraphics[width=8cm]{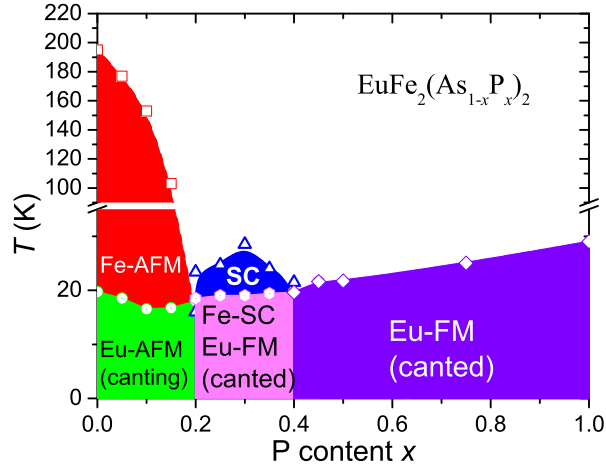}
\caption{Electronic phase diagram of
EuFe$_{2}$(As$_{1-x}$P$_x$)$_2$. Fe-AFM (Eu-AFM) denotes
antiferromagnetism in the Fe (Eu) sublattice; SC and FM refer to
superconductivity and ferromagnetism, respectively.}
\end{figure}

Compared with the "old" systems with the coexistence of SC and FM,
then, what is special in the present Eu122 system? Here we point out
three distinguishable features. 1) The FM has an unprecedentedly
large saturated moment ($\sim7 \mu_{B}$/Eu). 2) The magnetic
ordering temperature $T_M$ is very near the superconducting
transition one $T_{c}$, and both are relatively high. Once
$T_{c}<T_{M}^{Eu}$, SC disappears. 3) SC coexists with FM in broad
ranges of temperature and field. All these features make the Eu122
system worthwhile for further study.

Finally, we would like to discuss why SC and FM are compatible in
the Eu122 system. First of all, the superconducting upper critical
fields $H_{c2}$ in 122 Fe-based superconductors are very high
(\emph{e.g.}, $\sim$60 T for BaFe$_{2}$As$_{2}$\cite{yuan}). The
$H_{c2}$ values are actually higher than the hyperfine field on Eu
nucleus ($\sim$28 T) from the M\"{o}ssbauer
measurement\cite{felner-jop}. This makes SC survive even in the
presence of a strong internal field via RKKY interactions. Secondly,
it was indicated that all the five Fe-3$d$ orbitals contribute the
density of states near Fermi level.\cite{multiorbital} However, only
$d_{yz}$ and $d_{zx}$ orbitals are most probably related to
SC.\cite{sct} The $d_{x^{2}-y^{2}}$ and $d_{z^{2}}$ electrons are
supposed to be responsible for mediating RKKY
interactions.\cite{Ren-Ni} Consequently, both SC and FM are well
supported. In this sense, therefore, the Eu122 system seems to be a
very rare platform for studying the challenging issues of the
coexistence of SC and FM.

\section*{Acknowledgments}
We would like to thank I. Felner, J. H. Dai and H. Q. Yuan for
helpful discussions. This work is supported by the NSF of China
(Nos. 10934005 and 10931160425), National Basic Research Program of
China (No. 2010CB923003).

\end{document}